\documentclass[final,3p,times,twocolumn]{elsarticle}

\usepackage{graphicx,bm,epsf,float,amssymb,amsmath}

\journal{Nuclear Instruments and Methods}

\begin{document}

\begin{frontmatter}



\title{A Note on Neutron Capture Correlation Signals, Backgrounds, and Efficiencies}

\author[llnl]{N.~S.~Bowden\corref{cor1}}\ead{nbowden@llnl.gov}
\author[llnl]{M.~Sweany}
\author[llnl]{S.~Dazeley}
\cortext[cor1]{Corresponding Author. Tel.: +1 925 422 4923.}

\address[llnl]{Lawrence Livermore National Laboratory, Livermore, CA~94550, USA}

\begin{abstract}

A wide variety of detection applications exploit the timing correlations that result from the slowing and eventual capture of neutrons. These include capture-gated neutron spectrometry, multiple neutron counting for fissile material detection and identification, and antineutrino detection. There are several distinct processes that result in correlated signals in these applications. Depending on the application, one class of correlated events can be a background that is difficult to distinguish from the class that is of interest. Furthermore, the correlation timing distribution depends on the neutron capture agent and detector geometry. Here, we explain the important characteristics of the neutron capture timing distribution, making reference to simulations and data from a number of detectors currently in use or under development. We point out several features that may assist in background discrimination, and that must be carefully accounted for if accurate detection efficiencies are to be quoted.

\end{abstract}


\end{frontmatter}

\section{Introduction}
\label{sec:intro}


Neutron detection systems that incorporate a neutron capture agent or that produce a unique detector response to neutron capture have many applications. Some of these systems rely on timing correlations between a preceding interaction and a neutron capture to select events of interest.  In these cases, it is important to understand the physical mechanisms involved in determining the form of the timing correlation on which the event selection rests. This includes both the neutron production process, the neutron transport (including any signal it might produce), and the neutron capture itself. Furthermore, background processes that can produce similar or identical timing correlations must also be considered.

Recently, there has been considerable activity in producing and evaluating neutron capture correlation detectors. Examples of capture-gated neutron spectrometers~\cite{capture} include those for deep underground neutron background measurements~\cite{NIST}, nuclear physics measurements~\cite{N-Star}, and fissile material detection~\cite{pozzi_LGB,LGB3, pozzi_Cd}. Efforts to exploit correlations between neutron captures for neutron background measurements~\cite{soudan} and fissile material detection~\cite{sweany} are also underway. Finally, neutron capture correlations are central to reactor antineutrino detection, and therefore to the growing number of efforts to use this technology for reactor monitoring~\cite{SONGS1, Nucifier, SNL_LiZnS, Nelson_LGB, Suekane}. It is worth noting that these examples use a wide range of capture agents (e.g. Gd, $^6$Li, $^{113}$Cd, and $^{10}$B) and capture agent geometries (e.g. homogenous and inhomogeneous capture agent distributions).

Given these activities, we feel it is timely to review the primary signal and background processes for these applications, as well as the timing correlations that result. In particular, the timing distributions of correlated events depend strongly on the capture agent(s) and geometry used and can vary between signal and background processes. Predictions of detection efficiency must take these effects into account, and can potentially exploit them for background rejection. We will begin by reviewing the physical processes giving rise to correlation signals, followed by descriptions of the timing distributions that result from such processes and how these must be understood for accurate efficiencies to be calculated. We will then discuss how these and higher-order timing distributions can be exploited for background discrimination. 

In the following we will consistently refer to an ``event'' as a collection of distinct energy deposits in a detector that are associated with an initiating instantaneous physical process either external to the detector or in the detection medium, such as a fission or an antineutrino interaction.

\section{Correlated Neutron Production Processes}
\label{sec:production}

In this section we briefly review the important physical processes that can produce neutrons, and that can be identified and/or studied using neutron capture correlations. It is useful to distinguish two classes of neutron capture correlations:
\begin{itemize}
\item{ ``Prompt-Capture'' (PC) -- the time difference between an interaction occurring simultaneous with the initiating process and the capture of a neutron,}
\item{``Capture-Capture'' (CC) -- the time difference between the capture of two or more neutrons, where each neutron capture occurs sometime after the initiating process. }
\end{itemize}
We emphasize this distinction since, as will be discussed in Sec.~\ref{sec:TimingDists}, these event classes can produce different timing correlation distributions.

\subsection{Muon Spallation}
Direct muon spallation often produces multiple high energy neutrons, which can in turn initiate hadronic showers resulting in yet more neutrons.  Neutron production is both muon or neutron energy and medium dependent \cite{mun}. Neutron capture correlations can be measured between the initiating muon and a subsequent neutron capture (PC), or between the capture of any two produced neutrons (CC).  Furthermore, this process often produces showers with $>2$ neutrons, which can in turn result in the correlated detection of $>2$ neutron captures. 

\subsection{Spontaneous Fission}
\label{sec:SF}
Many actinides spontaneously fission. Important examples include $^{252}$Cf and $^{240}$Pu. The resulting simultaneous emission of multiple gamma rays and neutrons can provide a powerful means of detecting the presence of such material. Typical neutron multiplicities per fission are $\approx3$, and the neutron energy follows a power law fission spectrum ranging up to $\approx10$~MeV and mean energy $\approx1$~MeV that varies slightly with isotope. Depending on the details of the detection scheme, correlations can be observed between a prompt fission gamma-ray and a neutron capture (PC), a fast neutron recoil and a neutron capture (PC), or between multiple neutron captures (CC). We do not explicitly consider the possibility of fission chains here, but on the $\mu$s time scales being considered their net effect will be to increase the average neutron multiplicity.

\subsection{($\alpha$,n) Reactions}

A convenient means of producing a neutron source is to expose a Be or B target to $\alpha$ particles emitted by an actinide nucleus (e.g. $^{241}$Am). The resulting ($\alpha$,n) exchange reaction produces a neutron with an energy in the $0-10$~MeV range. The target daughter nucleus is often produced in an excited state which promptly decays via $\gamma$-ray emission. A common source of this type is an encapsulated mixture of $^{241}$Am and Be particles (an AmBe source). Many neutrons produced by a typical AmBe source are accompanied by a $4.4$~MeV $\gamma$-ray emitted by the daughter $^{12}$C nucleus. Depending on the details of the detection scheme, correlations can be observed between the prompt $\gamma$-ray and capture of the neutron or a fast neutron recoil and the capture of the neutron (both PC).

We note that an AmBe neutron calibration source can only produce PC neutron correlation signals, while a $^{252}$Cf source can produce both PC and CC signals (Sec.~\ref{sec:SF}).

\subsection{Antineutrino Interactions}

Electron antineutrinos can be detected via the inverse-beta interaction: $\bar{\nu}_e + p \rightarrow e^{+} + n$. Immediate detection of the final state positron, followed a short time later by detection of the capture of the neutron, forms a PC correlation. Due to the very small cross section for the antineutrino interaction, an intense source is required, e.g. a nuclear fission reactor.

\begin{figure}[tb]
\centering
\includegraphics*[width=0.45\textwidth]{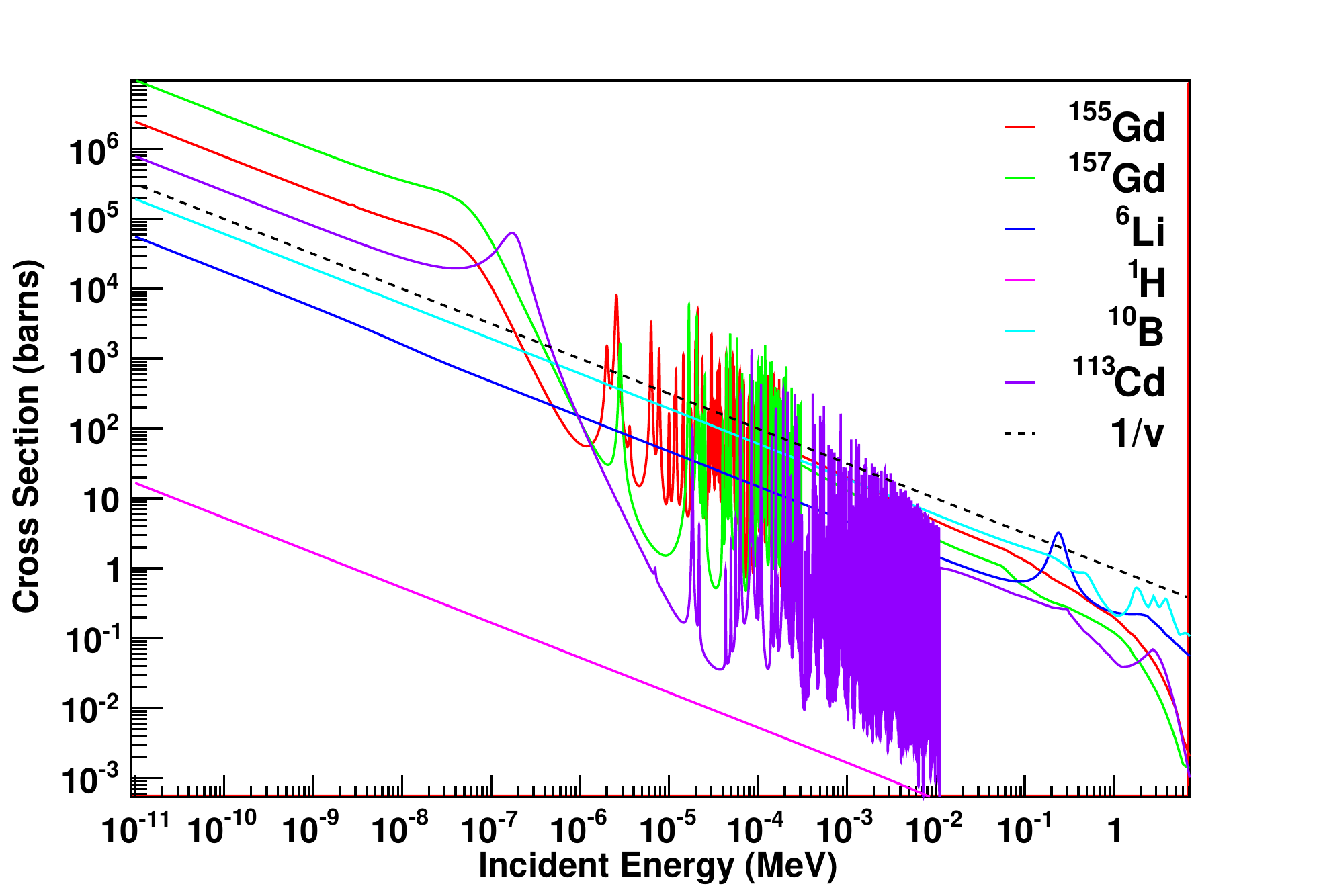}
\caption{The capture cross sections of the commonly used capture agents: $^{155}$Gd, $^{157}$Gd, $^6$Li, $^{113}$Cd, $^{10}$B and hydrogen. The energy dependencies of these cross-sections must be considered, particularly for Gd and non-homgeneous detector geometries.} \label{fig:xsections}
\end{figure}

\section{Neutron Capture Correlation Timing Distributions}
\label{sec:TimingDists}

\begin{figure}[tb]
\centering
\includegraphics*[width=0.45\textwidth]{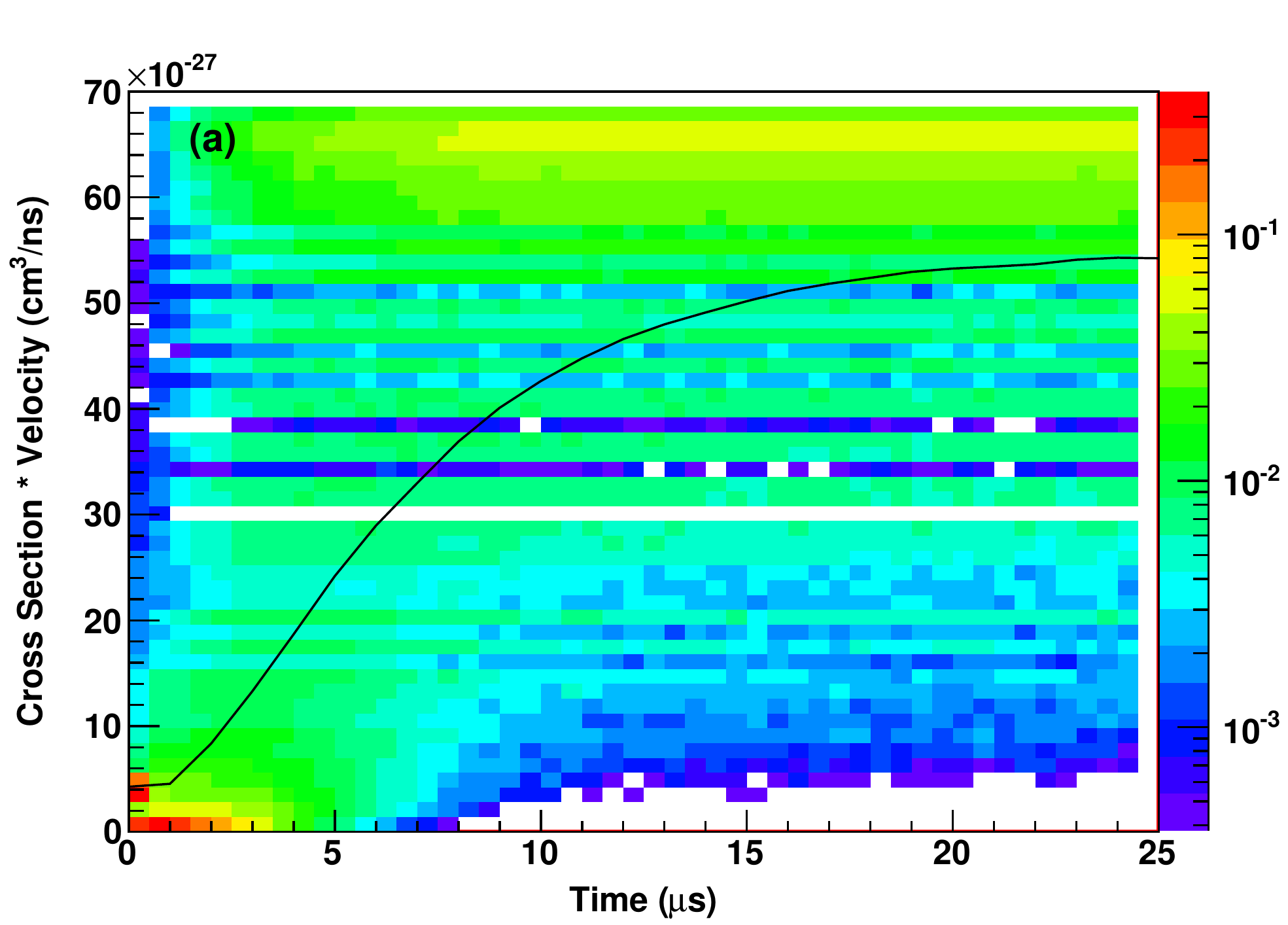}
\includegraphics*[width=0.45\textwidth]{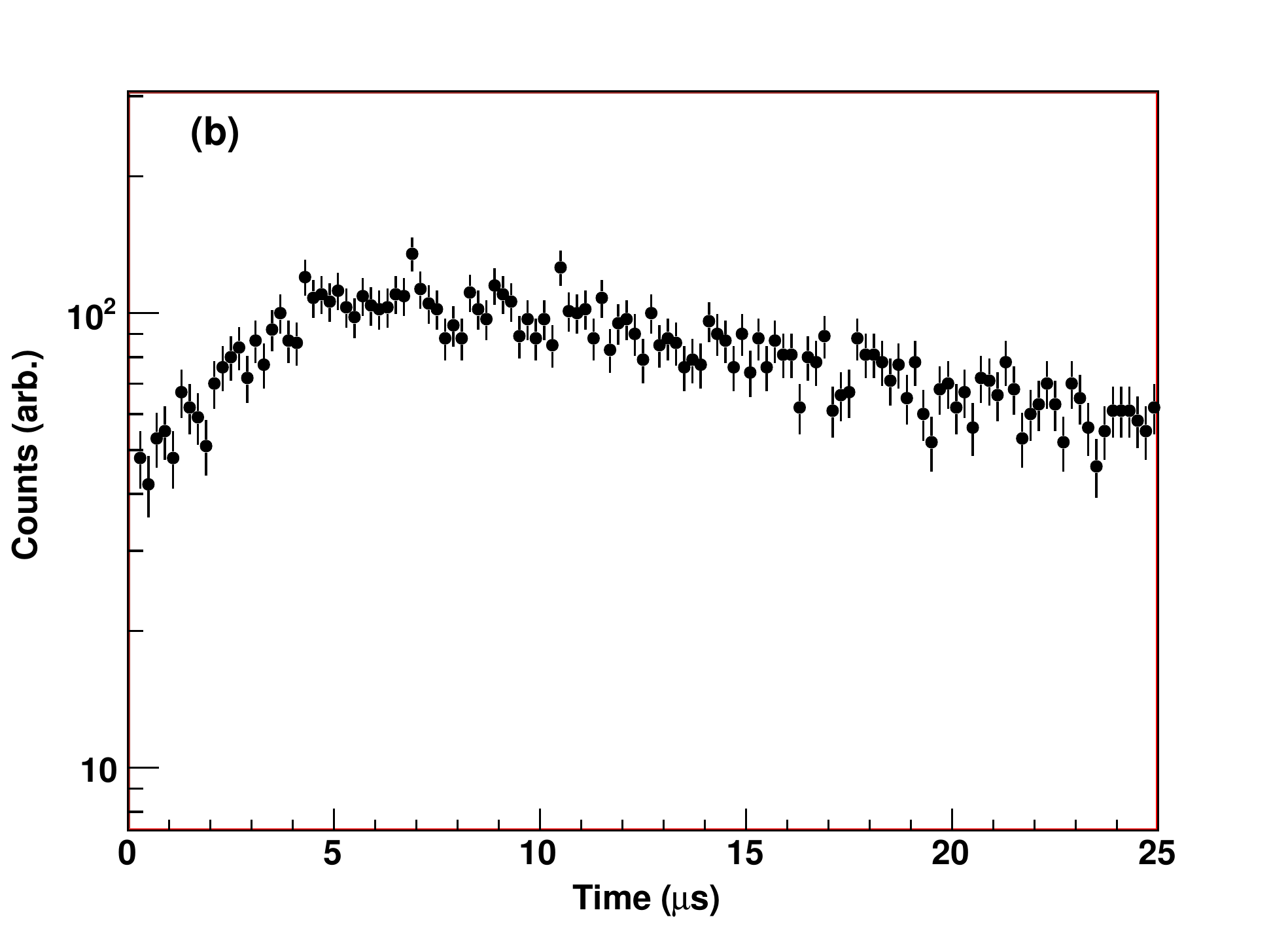}
\caption{(a) $P_{cap}$ as a function time for 0.1 MeV neutrons introduced to 0.01\% $^{155}$Gd doped water at $t=0$.  $P_{cap}$ increases to a constant value as the neutron is moderated. (b) The resulting capture time distribution. Once $P_{cap}$ reaches a constant value, the capture time distribution is an exponential function of time. } \label{fig:caprobGd}
\end{figure}

\begin{figure}[tb]
\centering
\includegraphics*[width=0.45\textwidth]{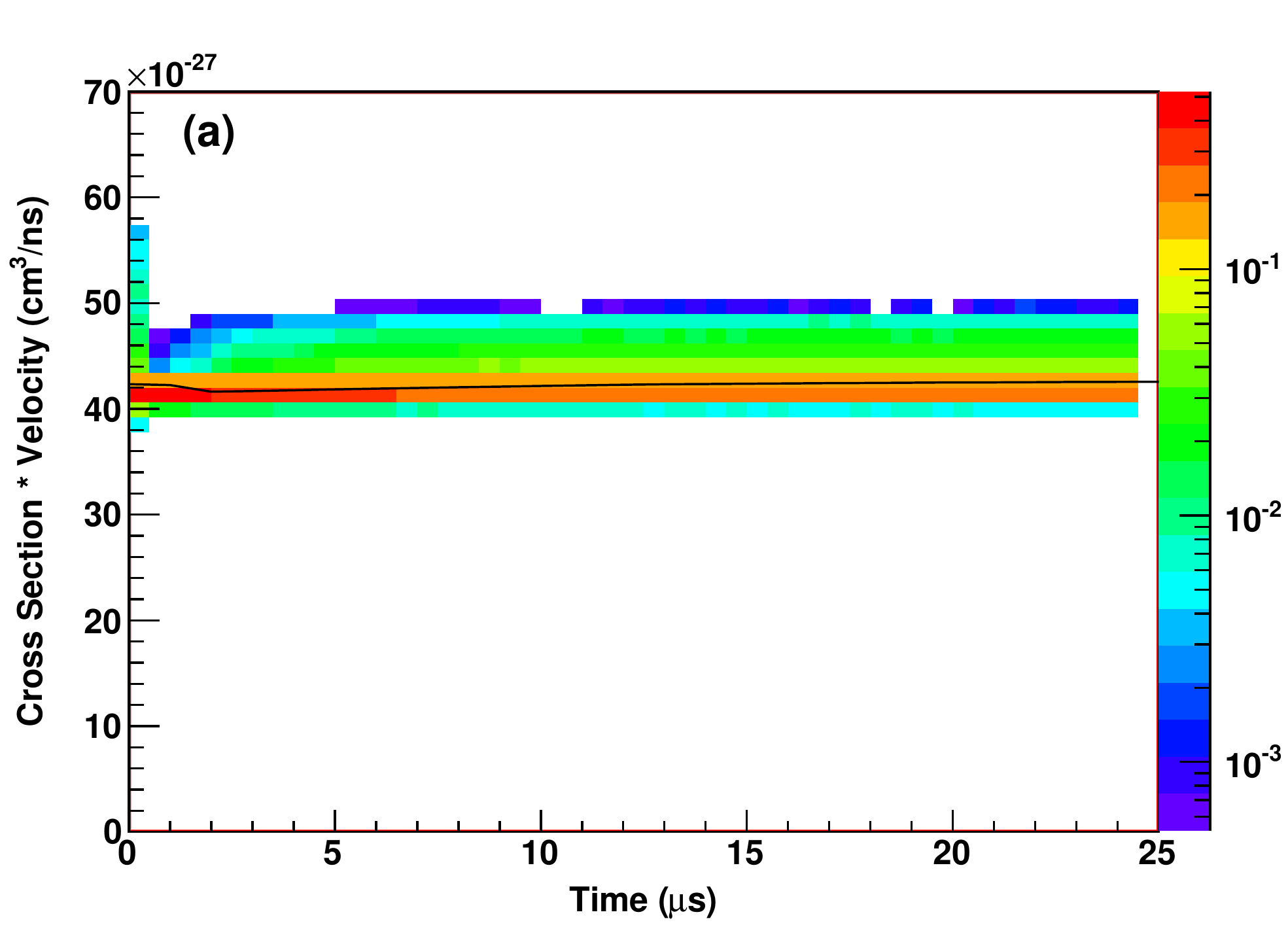}
\includegraphics*[width=0.45\textwidth]{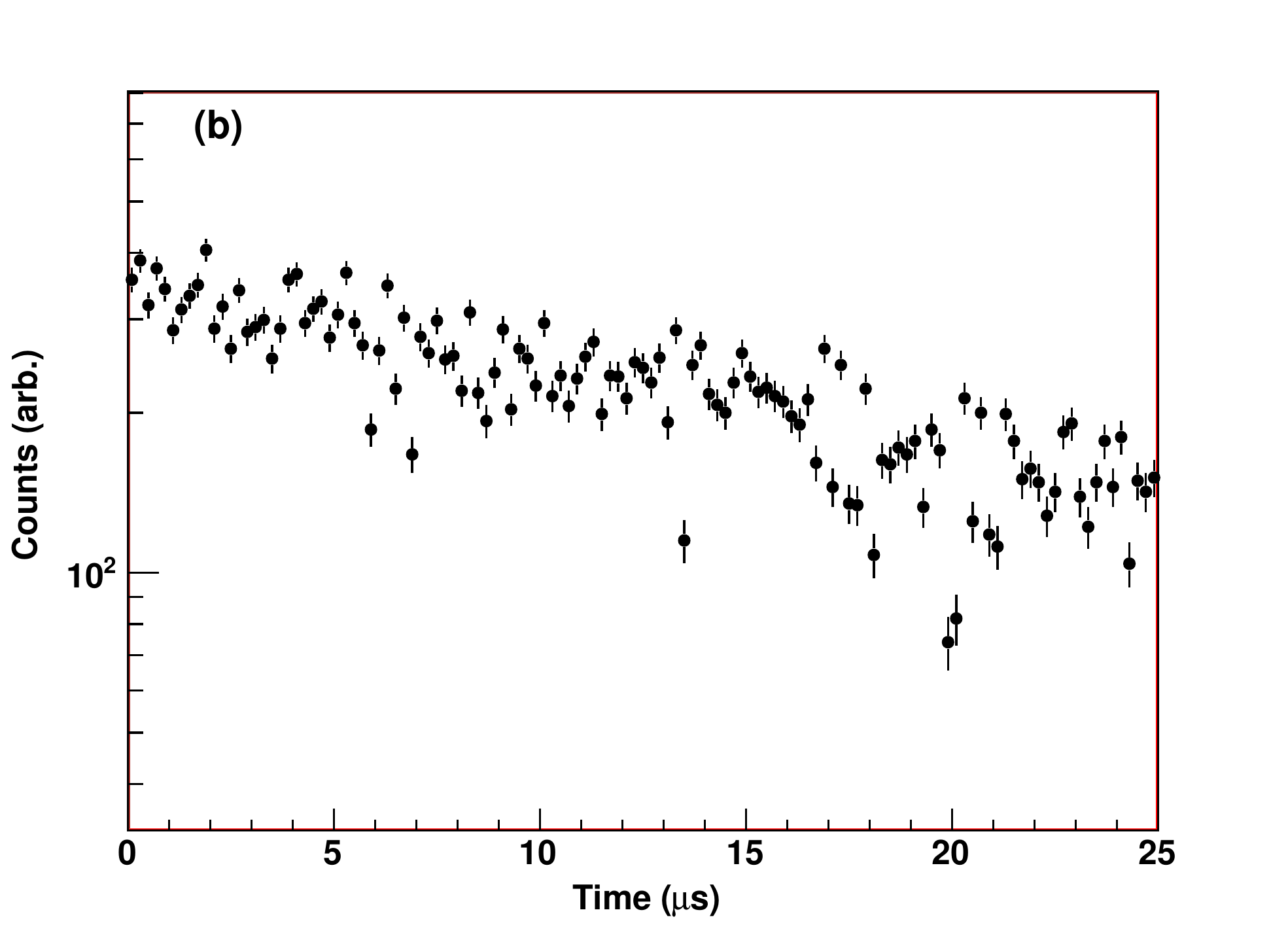}
\caption{(a) $P_{cap}$ as a function time for  0.1 MeV neutrons introduced to 0.4\% $^6$Li doped water at $t=0$.  $P_{cap}$ maintains a constant value independent of time (neutron energy). (b) The resulting capture time distribution. Since $P_{cap}$ is constant, the capture time distribution is an exponential function of time.} \label{fig:caprobLi}
\end{figure}

Experimental techniques that exploit neutron capture correlations typically measure the distribution of time intervals between some initial deposition and a neutron capture. The form of the timing distribution therefore depends on the relationship between that initial deposition and the neutron that captures, and the dynamics of the capture of that neutron. Let us first consider the neutron capture dynamics in the most simple situation, a homogenous detection medium into which a single non-relativistic neutron of energy $E$ is introduced. The probability that the neutron captures in a time interval $dt$ is proportional to: 
\begin{equation}
P_{cap}(E) \propto \sum_i  \sigma_i(E) w_i v dt,
\label{eq:cap_prob}
\end{equation}
where $v$ is the neutron velocity, the product $v dt$ represents the distance traveled by the neutron in the material in the interval $dt$, and the $\sigma_i$ and $w_i$ are the neutron capture cross-sections and stoichiometric fractions respectively of all constituents of the detection material. As the neutron energy changes due to collisions with the medium, so too will $P_{cap}$, i.e.:
\begin{equation}
P_{cap}(t) \propto \sum_i  \sigma_i(E(t)) w_i v(t) dt,
\label{eq:cap_prob_t}
\end{equation}
where the time dependence of $E$ and $v$ is explicitly noted. 

In the simplifying case where the capture cross-section is of the form $1/v$, $P_{cap}(E)$ is independent of energy and the capture time distribution is therefore simply an exponential function of  time. As can be seen in Figure~\ref{fig:xsections}, this is condition holds for H, $^6$Li and $^{10}$B below $\approx10$~keV. Otherwise, for isotopes whose capture cross-section does not follow that simple form, the capture timing distribution will depend on the initial energy of the neutron. For example, in the case of $^{155}$Gd, $^{157}$Gd and $^{113}$Cd, if the neutron has energy greater than $\approx 0.1$~eV, $P_{cap}$ will initially be relatively small, before increasing to a constant value as the neutron slows. 

These two situations are contrasted in Figures~\ref{fig:caprobGd}\&\ref{fig:caprobLi} which display $P_{cap}(E)$ as a function of time for $50,000$~simulated neutrons of $0.1$~MeV initial energy in homogeneously $^{155}$Gd and $^6$Li doped water respectively. The loadings (0.1\% b.w. $^{155}$Gd, 4.0\% b.w. $^6$Li) are chosen so that $P_{cap}$ is approximately equal at thermal neutron energies. Also shown is the average value of $P_{cap}(E)$ for each time step. The difference between $^6$Li and Gd doping is evident: for the former $P_{cap}$ is approximately constant, while for the latter $P_{cap}$ increases to a constant value as the neutron is  moderated, and the capture cross section increases, over time. As can be seen, the typical timescale for a neutron to reach the constant $P_{cap}$ regime for this Gd doping is $\approx 7~\mu$s.

\begin{figure}[tb]
\centering
\includegraphics*[width=0.45\textwidth]{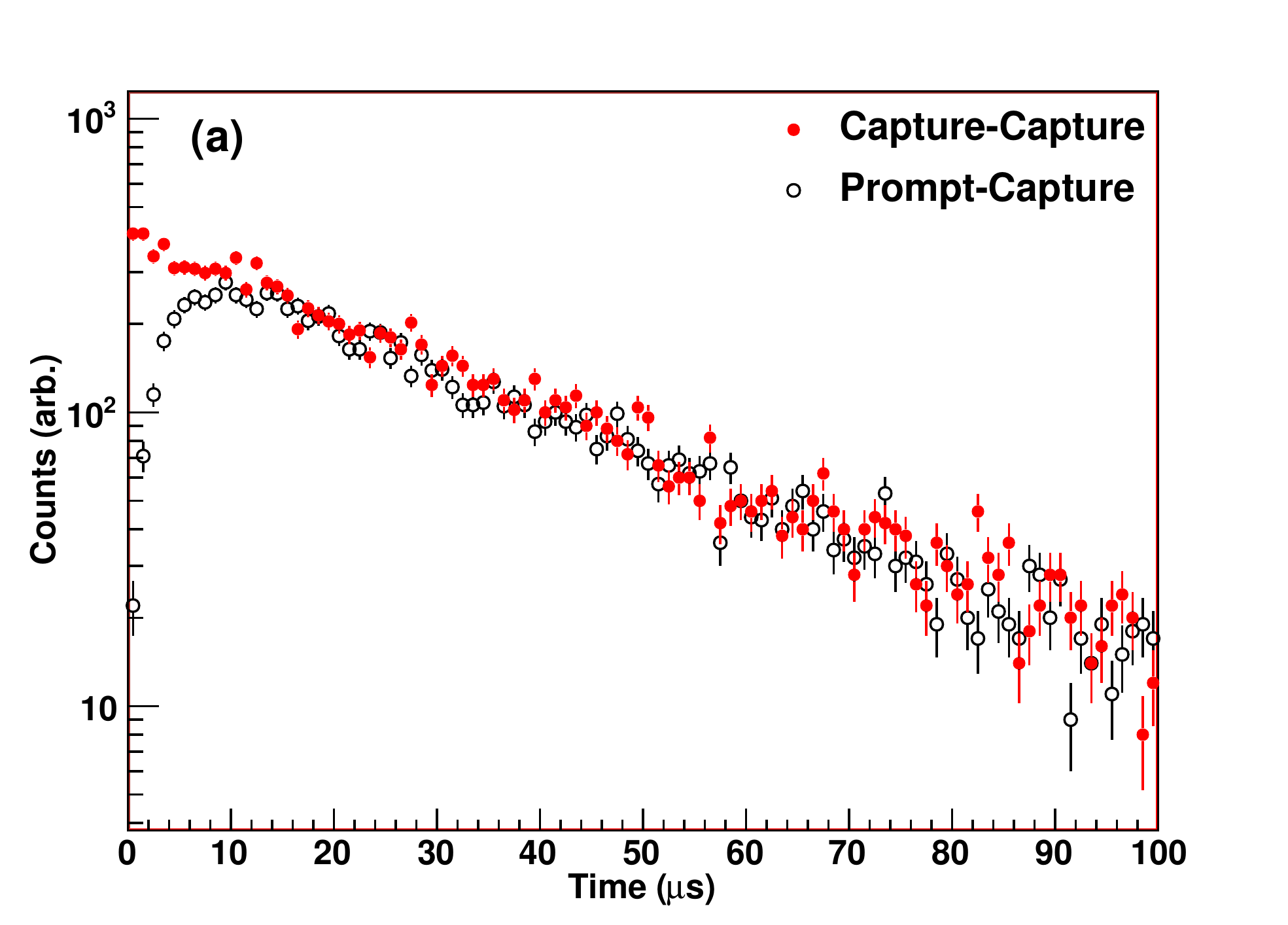}
\includegraphics*[width=0.45\textwidth]{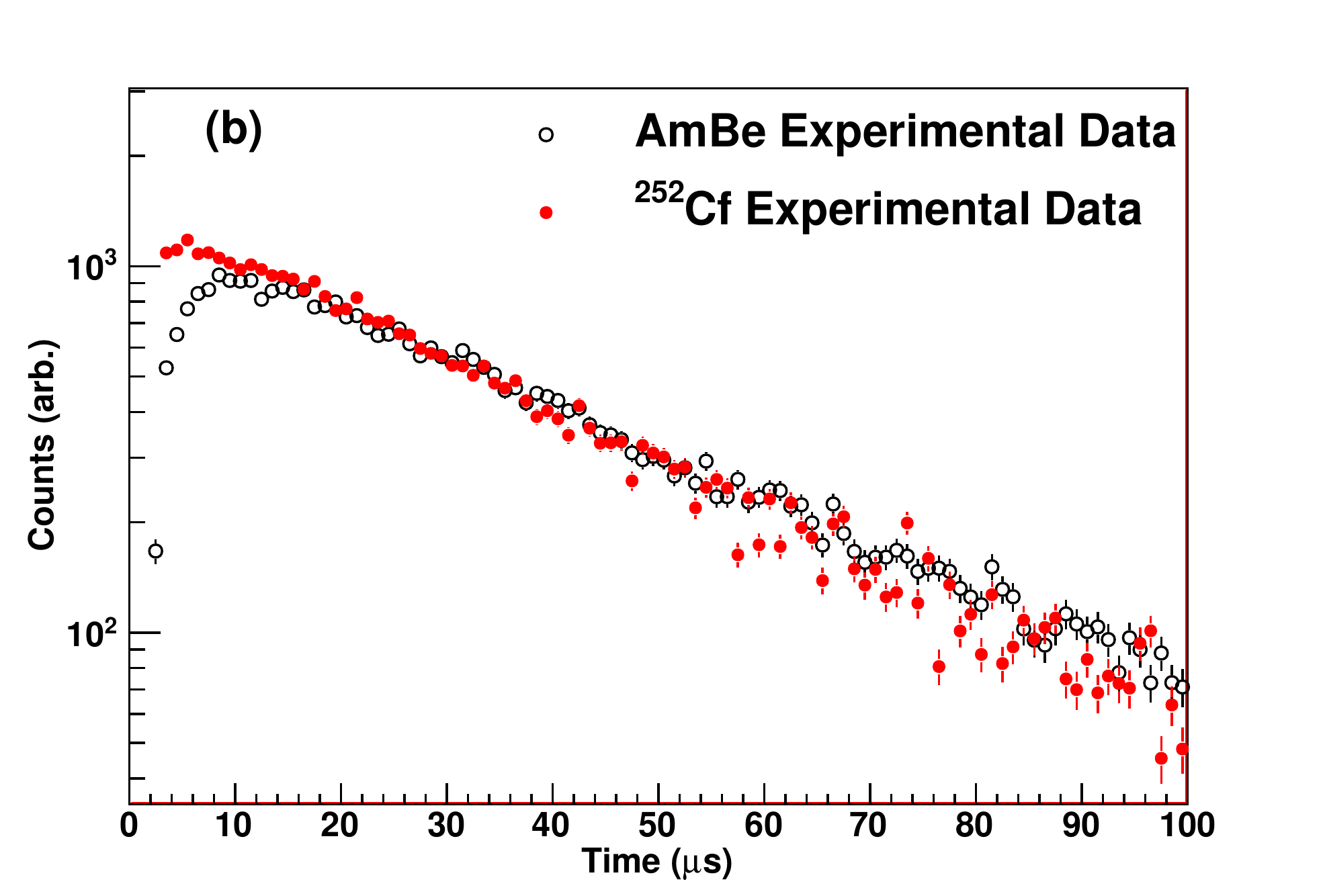}
\caption{Here we compare neutron capture time distributions for two classes of events: one where a neutron enters a Gd doped detector with $\approx1$~MeV energy at time=0 (PC, black); and where multiple neutrons enter the detector at unknown time and one records the time interval between their captures (CC, red). The low capture rate for PC events at short times reflects the lower Gd capture cross section at higher neutron energies. Panel (a) displays simulated data, while panel (b) displays experimental data.} \label{fig:capture_sim_data}
\end{figure}

This leads us to the reason for making the distinction between PC and CC events. Recall that the time interval measured for a PC event is that between a deposition occurring simultaneous with the initiating event and the capture of a neutron produced by the initiating event. Therefore, the timing distribution of PC events will depend upon the initial neutron energy, the capture agent and the detector geometry used. On the other hand, the time interval measured for a CC event is that between the capture of two or more neutrons produced simultaneously by the initiating event, where each neutron capture occurs sometime after that initiating event. The important point here is that, typically, both neutrons will have moderated before either captures. Therefore, even from $t=0$ (the time at which the first neutron captures) $P_{cap}$ will typically be constant, and the capture time distribution for CC events will be a simple exponential even at short times for a homogeneous detection medium.

This effect is demonstrated in Figure~\ref{fig:capture_sim_data} for both simulated and experimental data. Here,  timing distributions for PC and CC events measured in the Gd doped Water Cherenkov detector described in~\cite{sweany} are compared. A PC sample is collected using an AmBe neutron source, while the CC sample is from a $^{252}$Cf neutron source; although the distribution is not purely CC, the PC fraction is small.  The expected difference in capture rate at short times is evident.

It is difficult to extend the above discussion in a general way to detector systems that incorporate multiple and/or inhomogeneously distributed capture agents. Similar considerations regarding the energy dependence of the capture cross-sections will apply, with the additional complication that the neutron will occupy un-doped material for much of the time. Monte-Carlo simulation tools are indispensable in the design of and interpretation of data from such systems. Validation of those simulations with well understood neutron sources is essential. 

\begin{figure}[tb]
\centering
\includegraphics*[width=0.45\textwidth]{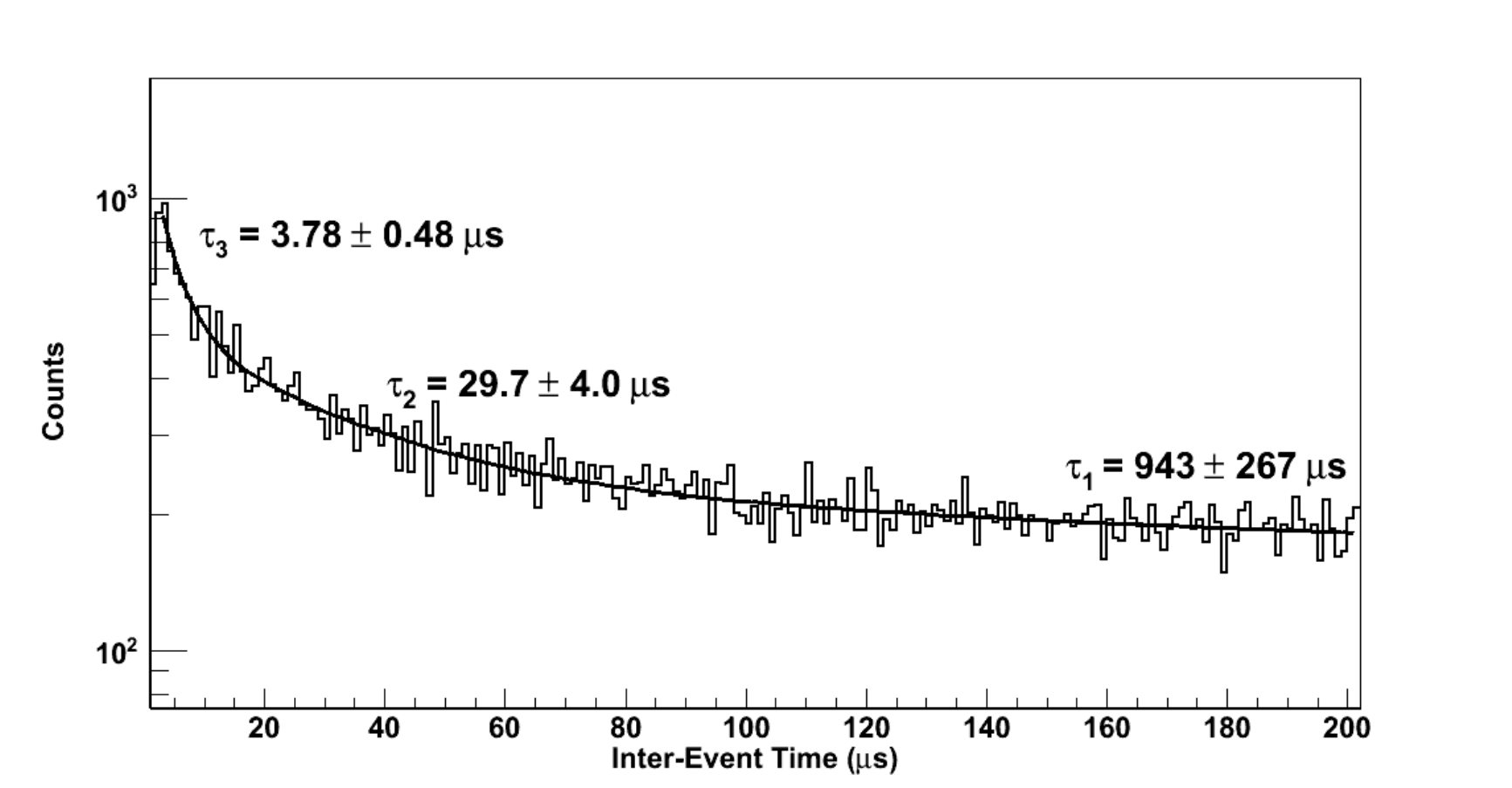}
\caption{A multi-component neutron capture timing distribution resulting from complex inhomogenous detector geometry. Figure~23 from~\cite{Nelson_thesis}.} \label{fig:LGB}
\end{figure}

By way of an example consider Fig.~\ref{fig:LGB}, which displays the neutron capture correlation timing distribution of the inhomogenous, multiple capture agent detector described in~\cite{Nelson_LGB} in response to a $^{252}$Cf source. Two distinct capture time constants are observed (the exponential feature with $\tau \approx 900\mu$s is due to the random coincidence of singles). Investigations using a Geant4 simulation suggest that these features depend both on the degree of inhomogenaity and the different energy dependence of the capture agents used ($^6$Li and $^{10}$B vs Gd)~\cite{Nelson_thesis}.

Finally, we note a feature of the neutron capture timing distributions due to sources that produce two or more neutrons. The discussion above focussed upon the situation where a single neutron is introduced. When two or more neutrons are introduced simultaneously, e.g.~due to a spontaneous fission or muon spallation event, Eq.~\ref{eq:cap_prob_t} holds for each neutron independently. Therefore, the probability of any neutron capturing in an interval $dt$ when there are $N$ neutrons present is:
\begin{equation}
P_{cap}(t) \propto \sum^N_i \sum_j \sigma_i(E_i(t)) w_j v_i(t) dt,
\label{eq:cap_prob_t_N}
\end{equation}
where $E_i$ and $v_i$ are the energy and velocity of the $i$th neutron. To a reasonable approximation, the average value of $P_{cap}(t)$ will be:
\begin{equation}
\bar{P}_{cap}(t) \propto N \sum_i \sigma_i(\bar{E}(t)) w_i \bar{v}(t) dt,
\label{eq:cap_prob_t_N_avg}
\end{equation}
where $\bar{E}(t)$ and $\bar{v}(t)$ are the average neutron energy and velocity at time $t$. That is, the probability of a neutron capture occurring increases by a factor of $N$, and subsequently the time interval between successive captures will decrease by that same factor. Therefore, the total measured neutron capture timing distribution measured from a source that produces multiple neutrons will be a sum of the distributions for $1...N$ neutrons, each weighted by a factor determined by the source neutron multiplicity distribution and the neutron detection efficiency of the detector.

\begin{figure}[tb]
\centering
\includegraphics*[width=0.45\textwidth]{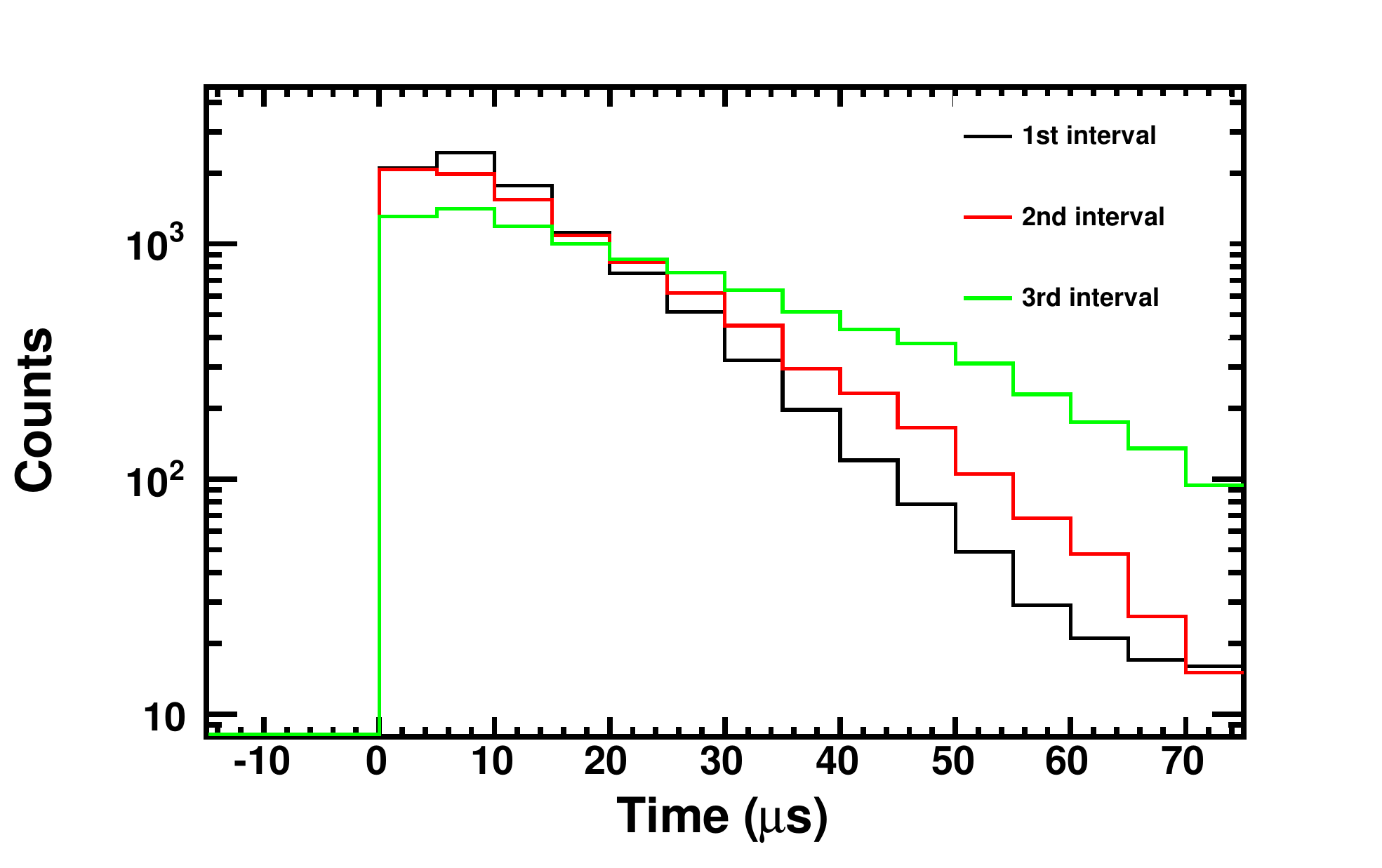}
\caption{The successive intervals between neutron captures, where 3 neutrons are initially present. Fitting the simple exponential function $e^{t/\tau}$, yields $\tau=30.1~\mu$s, $16.0~\mu$s, and $11.9~\mu$s for the 3rd (green), 2nd (red), and 1st (black) intervals respectively. Taking the ratio of each $\tau$ to that for the 3rd interval yields $0.39:0.55:1$, close to the expected $\frac{1}{3}:\frac{1}{2}:1$.} \label{fig:MultiNeutron}
\end{figure}

This effect is illustrated in Fig.~\ref{fig:MultiNeutron} which shows experimental muon spallation data from a detector similar to that described in~\cite{sweany}. Here, closely spaced sequences of 4 depositions consistent with neutron capture have been selected, so that at the time the first deposition occurs there are $3$~neutrons present in the detector. Specifically, the interval between the first and fourth deposition is required to be less than $200~\mu$s. To ensure that the selected sequence is not a subset of a longer sequence, an additional ``isolation'' selection is applied: the first and fourth depositions must be at least $100~\mu$s from the preceding and subsequently depositions, respectively. One can see that the successive intervals between these depositions increase. The ratio of the capture time constants relative to that for the last interval very nearly follows the expected $\frac{1}{3}:\frac{1}{2}:1$ pattern. 

\section{General Considerations for Detection Efficiency Calculations}
\label{sec:eff}
The preceding discussion demonstrates how the neutron capture timing distribution can depend upon the event type, detector geometry, neutron capture agent, and neutron source. To state the obvious, calculations of detection efficiency must therefore take all of these factors into account. In practice, this requires careful simulation of the particular detector configuration begin considered, for the event type of interest. Care should be taken to validate that simulation against data with a known event type. We note that judicious use of $(\alpha$,n) and Spontaneous Fission (SF) sources provides a means of measuring the detector response to pure PC and CC event samples. 

Since an $(\alpha$,n) source like AmBe produces only a single neutron, it provides a pure PC sample that can be used for direct measurement of the detector response to this event class. The prompt signal in this case can be provided either by a proton recoil signal in the correlation detector or by the interaction of the de-excitation $\gamma$-ray often associated with $(\alpha$,n) reaction. A particularly clean method for this type of calibration was employed in~\cite{sweany}: detection of the de-excitation $\gamma$-ray in a separate detector was used not only to measure the neutron capture time distribution but also to estimate the absolute neutron capture efficiency.

Similarly, the multiple neutrons produced by a SF or muon spallation source can be used to obtain a CC event sample. The raw neutron capture timing distribution from such a source is an admixture of PC and CC events, since prompt $\gamma$-rays are produced by the fission and the fast neutrons released can produce prompt proton recoils. However, as was done in Fig.~\ref{fig:MultiNeutron}, closer examination of event sequences with three or more energy depositions closely spaced in time can provide a pure CC event sample. Since the last pair of depositions in such a sequence must both be due to neutron captures, the distribution of times between this last pair of the sequence will be that for CC events involving just a neutron pair.

\section{Potential Background Discrimination Strategies}

As discussed in Sec.~\ref{sec:production}, there are a wide range of processes that give rise to neutron capture correlation events. The physical process of interest to one application might very well be a troublesome background for another. For example, for capture-gated neutron spectrometry and antineutrino detection the signal of interest is always of type PC and is always the correlation of only two successive events (proton recoil followed by neutron capture for the first, positron followed by neutron capture for the second). Any multiple neutron capture sequences constitute background for these applications. The preceding discussions suggest a handful of circumstances in which observable differences in event classes can be used as a means of background discrimination. 

Most obviously, the difference in the capture time distribution at short times between PC and CC events for non-$1/v$~capture cross sections could be exploited via a simple timing selection, if the signal of interest produces a PC event. 

Additionally, we suggest a means to select or reject the multiple neutron capture sequences that can be produced by SF and, especially, muon spallation. Multiple neutron sequences can be readily identified by examining higher order correlation timing distributions. Indeed, for most applications that seek to study correlated pairs, examination of triple correlations suffices to reject much of the multiple neutron background. 

\begin{figure}[tb]
\centering
\includegraphics*[width=0.45\textwidth]{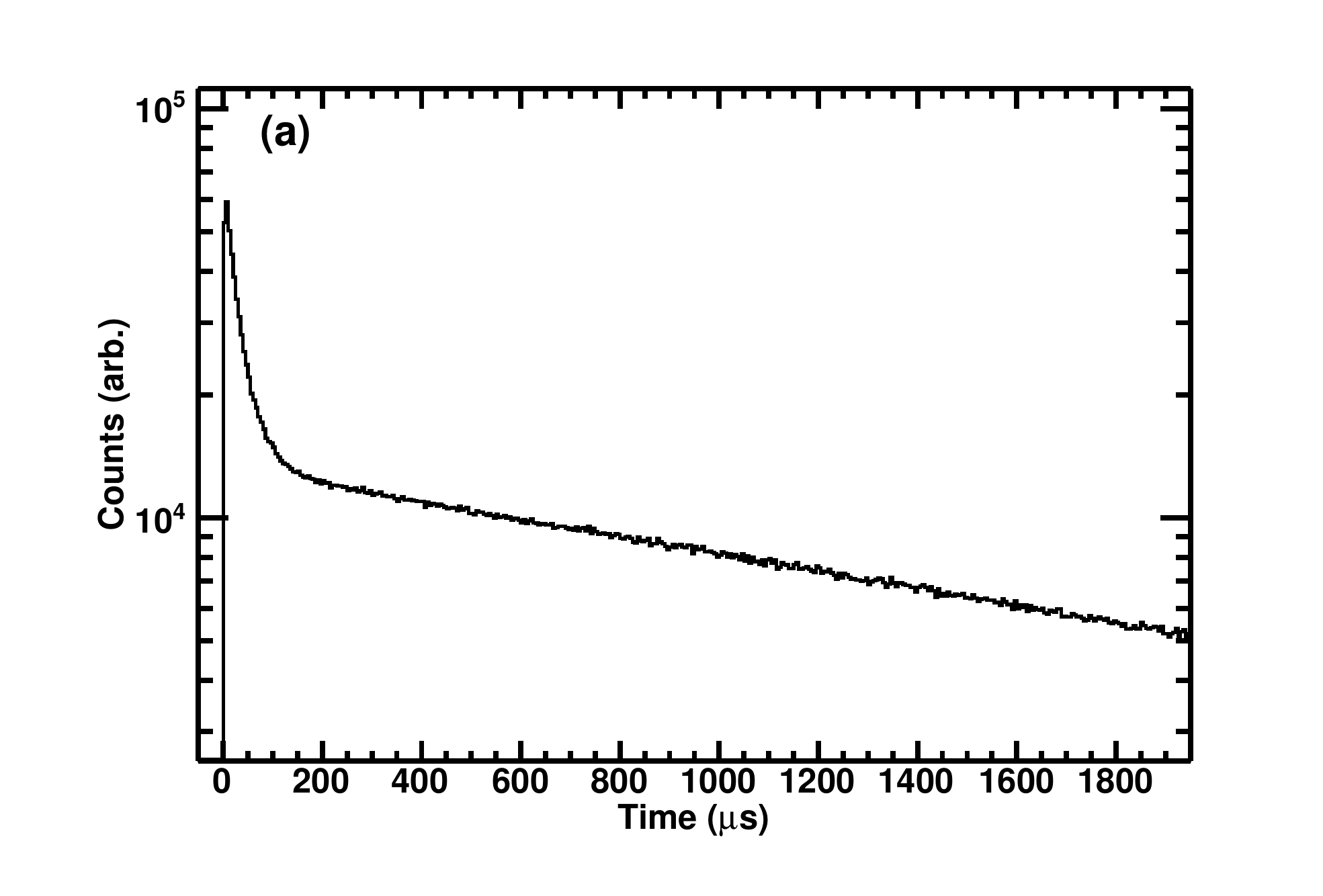}
\includegraphics*[width=0.45\textwidth]{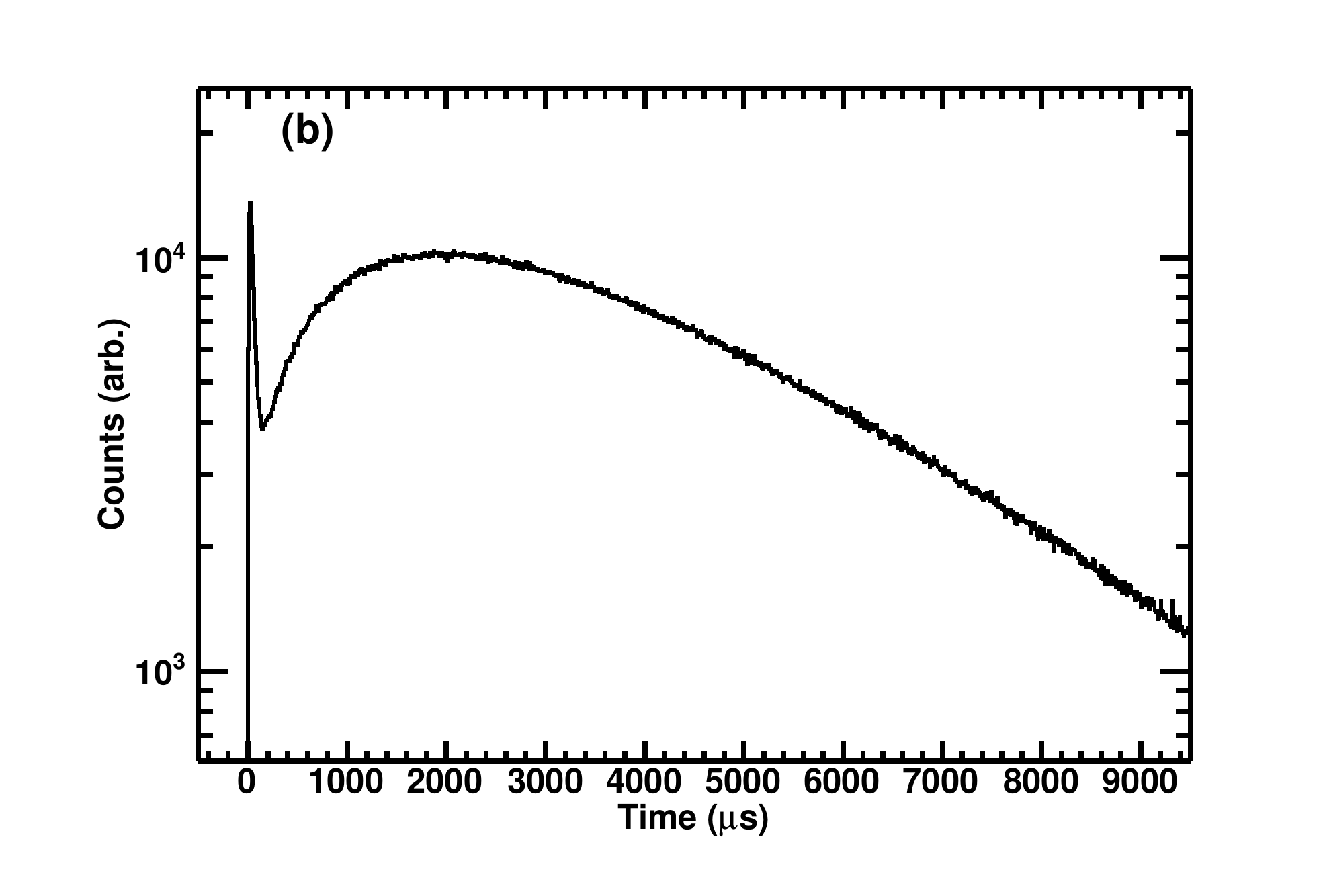}
\caption{Here we compare neutron capture time distributions for (a) double and (b) triple deposition intervals. The double interval displays two prominent exponential features, one due to correlated events and the other due to random coincidences. The triple interval reveals a short time  feature due to multiple neutron correlations and a broad distribution at larger times due to random coincidences.} \label{fig:triple_data}
\end{figure}

Specifically, examination of the interval between three successive depositions, in addition to the interval between pairs of depositions considered so far, is recommended. By way of illustration, Fig.~\ref{fig:triple_data} compares double and triple interval distributions from muon spallation events in a detector similar to that described in~\cite{sweany}. There are two evident features in both the double and triple interval distributions: one at short times due to correlated events and one at longer times due to the random coincidence of uncorrelated singles. The correlated (short time) feature in the triple interval distribution is due to correlation amongst three successive depositions, at least the last two of which are neutron captures. A selection cut rejecting event sequences with a short triple interval can therefore be effective at rejecting such higher multiplicity occurrences.

\section{Conclusion}
\label{sec:conclusion}

There are many detection applications that use neutron capture correlations to distinguish between the primary signal of interest and background processes.  These applications include, but are not limited to, capture-gated neutron spectrometers for underground neutron measurements and fissile material detection, and antineutrino detection for reactor monitoring.  Depending on the signal and background processes and the details of the detector design, the timing distribution between subsequent interactions in an event can vary considerably.   

Here, we have highlighted many of the important features of these timing distributions. The general point we wish to convey is that to develop a good understanding of detector efficiency, one must carefully study the response of the detector to the process of interest, as well as relevant background processes. While simulations are indispensable, we suggest validation of those simulations using specific neutron sources. Finally, we note that the variation in timing distributions for various processes can, in some cases, be exploited for background rejection.

\section*{Acknowledgements}

LLNL-JRNL-526291. We gratefully acknowledge support from the LLNL Laboratory Directed Research and Development program. This work was performed under the auspices of the U.S. Department of Energy by Lawrence Livermore National Laboratory under Contract DE-AC52-07NA27344.

\end{document}